# Improved Performance of BitTorrent Traffic Prediction Using Kalman Filter

**Hari Sankar S[1],    Pranav Channakeshava[2]**    and    **Pathipati Srihari[3]**

1,2 *Undergraduate, Department of Electronics and Communication Engineering, National Institute of Technology Karnataka, Surathkal, Mangalore 575025, India. Email:harisankarhss@gmail.com[1], pranav_chanakeshava@ieee.org [2]*
3  *Assistant Professor, Department of Electronics and Communication Engineering, National Institute of Technology Karnataka, Surathkal, Mangalore 575025, India. Email: srihari.js@gmail.com*

**ABSTRACT:** Supervising internet traffic is essential for any Internet Service Provider (ISP) to dynamically allocate bandwidth in an optimized manner. BitTorrent is a well-known peer-to-peer file-sharing protocol for bulky file transfer. Its extensive bandwidth consumption affects the Quality of Service (QoS) and causes latency to other applications. There is a strong requirement to predict the BitTorrent traffic to improve the QoS. In this paper, we propose a Kalman filter (KF) based method to predict the network traffic for various traffic data sets. The observed performance of KF is superior in terms of both Mean Squared Error (MSE) and total computation time, when compared to Auto Regressive Moving Average (ARMA) model.

**INTRODUCTION**
BitTorrent has emerged as a widely popular peer-to-peer (P2P) content distribution protocol over the Internet. It offers scalability and effective utilisation of bandwidth through swarming file delivery, which makes it very attractive for large file transfers. As opposed to the traditional client-side paradigm, BitTorrent uses a node in the network to share its content to its peers by acting as a source. Its "rarest first" approach for transfer of blocks of data emphasises on fast and efficient content distribution by organizing peers which share the same file, into a P2P network.

As P2P systems support important and reliable services like bulky data sharing, video streaming and Voice-over-IP, they have been reported to generate around 70% of internet traffic globally [11]. Their popularity has led to financial benefits for Internet Service Providers (ISPs), as users tend to upgrade to broadband for better performance. Given the high bandwidth utilization by the BitTorrent protocol, several issues of traffic engineering challenges have been continuously raised. The protocol's decisions are based on the application layer routing and are independent of the network topology, which causes an increase in the cross-ISP traffic [3]. As it generates a high volume of packet transaction, it causes users, who are not using the BitTorrent protocol, to experience packet loss if they are connected to the same LAN as that used by the BitTorrent users. This has led to ISPs limiting a user's P2P traffic to ensure fairness among all users of the network.

A lot of research has been done in the past, on looking into how this issue can be solved through ISPs analysing the traffic patterns, so that they can plan a throttling mechanism to ensure that all users get similar network resources. Hoong et al. (2012) proposed a prediction model using the Auto Regressive Moving Average (ARMA) time series model. In this paper, we propose an improved model for predicting the BitTorrent traffic based on Kalman Filtering which gives better accuracy, as measured by the mean squared error and is also faster in terms of the computation time. We compare the results of the ARMA prediction with that of the KF predictor to analyse their performance for seasonal BitTorrent network traffic data.

## BASICS
### BitTorrent Protocol

BitTorrent adopts a swarming approach through a set of peers, who can upload or download data from each other simultaneously. The protocol organizes the peers into a 'Torrent' network that has interactions among the peers and occasionally with the server for locating new peers. The peers are classified as a downloader or a seed, where downloaders are the peers who possess a part of the file while the seeds are the peers having the complete file. The seeds stay online in the network in order to share their data content while the downloaders obtain the parts of the file, which they do not have. The downloaders can subsequently become the seeds once they have the complete file.

In order to allow the client to arrange the pieces that are downloaded non-sequentially, each part of the file is hashed by a cryptographic torrent descriptor. As a result, the protocol allows for sudden halts in the download process at any instance without losing the already downloaded data. This also allows the client to search for the readily available pieces and initiate their download immediately rather than wait for sequential pieces that may not be available. These advantages play a crucial role in reducing the file transfer time.

The peer selection mechanism used by BitTorrent consists of the tit-for-tat, anti-snubbing, optimistic unchoking and upload-only strategies. These ensure that the download performance of the peers, who act as seeds in the network, is better than that of peers who act as free riders and do not contribute to the sharing [9].

### Auto Regressive Moving Average

Auto Regressive Moving Average (ARMA) is a time series stochastic model consisting of the Auto Regressive (AR) part which sums the previous observations, and the Moving Average (MA) part which consists of the sum of previous error terms. The model is represented by Eq. 1.

$$X_t = \sum_{i=1}^{p} \theta_i X_{t-1} + \sum_{i=1}^{q} \phi_i \varepsilon_{t-1} + \varepsilon_t \quad (1)$$

The first term in Eq. 1 refers to AR process and second term to MA process with their respective parameters $\theta_i$ and $\phi_i$. The terms are summed with Gaussian white noise $\varepsilon_t$. The parameters have to be determined to characterize a model. The p and q parameters characterise the AR and MA components of ARMA respectively. Different orders of ARMA are used to describe stationary time series and prediction based on the causal data.

### Kalman Filter

Kalman filtering (KF) is an established technique proposed by Kalman (1960). It is often used in estimation theory as an optimal predictor of the parameters of the state equations. The filter is fed the differences between the actual and predicted state variables as input. It minimizes the MSE and produces an estimate for the next time instance. The recursive equations process the newly observed data and estimate the state vector with the given information about previous data samples at time $t$.

KF assumes a linear stochastic equation for the estimated value $x_k$, which is a linear combination of its previous values plus the control signal $u_k$ and a process noise $w_k$, which is Gaussian. The measurement value $z_k$ is expressed as a linear combination of the signal value and the measurement noise.

$$x_k = A x_{k-1} + B u_k + w_k \quad (2)$$
$$z_k = H x_k + v_k \quad (3)$$

KF uses a prediction step which is given by the following time update equations:

$$\hat{x}_k = A \hat{x}_{k-1} + B u_k \quad (4)$$
$$P_k = A P_{k-1} A^T + Q \quad (5)$$

The equations project the next state and the error covariance. The prior estimate $\hat{x}_k$ gives the rough estimate before the measurement update correction. The prior error covariance is $P_k$. The prior values are used in the following measurement update equations.

$$K_k = P_k H^T (H P_k H^T + R)^{-1} \quad (6)$$
$$\hat{x}_k = \hat{x}_k + K_k(z_k = H\hat{x}_k) \quad (7)$$
$$P_k = (I - K_k H) P_k \quad (8)$$

These equations estimate x at time k by evaluating $\hat{x}_k$. Also, computing the error covariance $P_k$ is necessary for the future estimate, together with $\hat{x}_k$. Once the correction is done for the measurement update, the next prediction is estimated using the time update equations and this process is iterated.

KF prediction is recursive where it processes a data sample at a time instead of all the previous samples. The previous data is represented in $X_n$ and therefore, there is no need to store the data. The Kalman gain factor $K_k$ considers the previous estimates while predicting the future values.

**IMPLEMENTATION**
We base our study on the BitTorrent data collected using Wireshark, which is a network protocol analyser. A BitTorrent application was utilized to start the P2P communication and 5 sets of data packets were gathered during various times of the day. Network traffic was filtered for only TCP and UDP packets. We then convert the time series into CSV format and analyse it using MATLAB. We difference the time series to obtain the number of packets per second (frequency). Log transformation was applied to enhance the minute variations. We then perform a box centering using a rectangular window of size 10 and 50% overlap, to get a stationary time series.

We use the data sets for the prediction models. We compare the performance of the Kalman prediction with the ARMA model, by computing the Mean Squared Error (MSE) and the computation time. For the ARMA (p,q) model, different values of p and q were estimated and tried on all the 5 data sets. MSE is calculated based on the difference between the predicted data and actual data. The scaled stationary time series is also used for the Kalman predictor, to obtain its prediction accuracy. We assume a Gaussian random noise of variance 0.01.

**RESULTS AND DISCUSSION**
The prediction has been tested for 5 different data sets. For ARMA, we observe that ARMA (2, 1) predicts with the least MSE although ARMA (3, 0) is a close competitor. By adopting Kalman prediction, we find the prediction to be much improved. Figure 1 shows the superimposition of ARMA (2, 1) and KF for time series A which is validated against the actual bit torrent data. The comparison of MSE of KF prediction and different ARMA models for 5 data sets is shown in Table 1.

We have collected seasonal BitTorrent data and the MA parameter q is significant for the data since we assign same weights to all the observations irrespective of the peaks and valleys in the plot. ARMA (2, 1) proves to be better for predicting seasonal traffic patterns while ARMA (3, 0) is suitable for cyclic patterns which can be studied at a later stage. Variation in seasonal patterns allows us to estimate the pattern accurately only to a certain extent. This variation can be better estimated through the time update iterations of the Kalman filter and as such, a lower MSE is observed.

As seen from Table 1, it is clear that KF based method reduces the MSE by a factor of 5, and as such provides a more accurate model for predicting BitTorrent traffic. Table 2 compares the computation time required by the ARMA model and KF predictor as measured on MATLAB 2015a on a Windows 8 platform with 8GB RAM. It is evident that KF is much faster than ARMA estimation in terms of its speed of computation.

**CONCLUSIONS**
Given BitTorrent's popularity for large file transfer over the Internet, its use has been increasing significantly over the last decade. Though it represents an efficient mechanism for fast data transfer, its high bandwidth usage causing unequal resource distribution amongst network users has been an issue for ISPs, leading to restrictions placed on such transfers. One possible way to address this issue is through predictive measures by ISPs to ensure an equitable distribution.

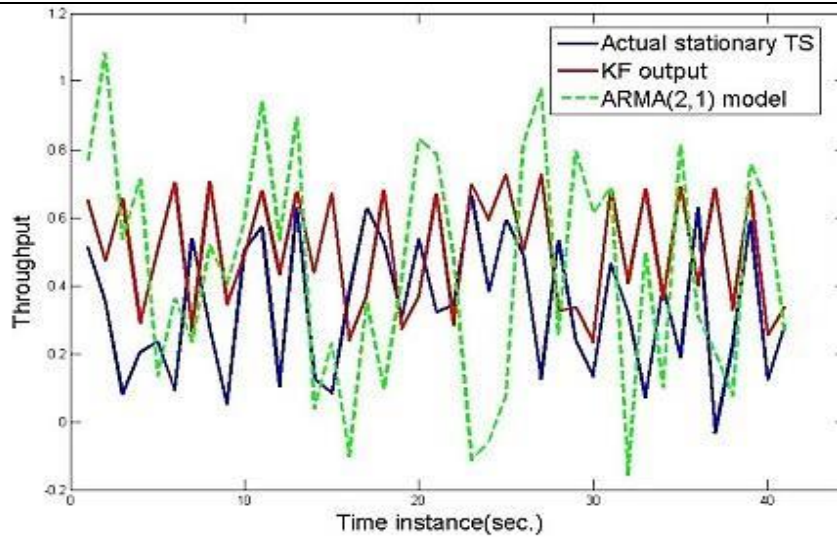

**Fig.1 BitTorrent Prediction using Kalman Filter predictor and ARMA (2,1)**

We propose the Kalman Filter based predictor as an effective model to estimate bandwidth requirements for BitTorrent traffic data. The technique proves to be a more accurate model as it achieves lower MSE when compared to previously proposed ARMA model by Hoong et al. (2012). Our prediction technique produces better estimates and is also superior in terms of its computation speed when compared to ARMA.

Thus, the prediction model can be used by ISPs to better understand BitTorrent traffic data, which will enable them to effectively allocate network resources in a more equitable manner amongst its users.

**Table 1 Comparison of MSE**

| SET | ARMA (2,0) | ARMA (2,1) | ARMA (2,2) | ARMA (3,0) | ARMA (3,1) | KF |
|---|---|---|---|---|---|---|
| A | 0.10 | 0.089 | 0.091 | 0.092 | 0.093 | 0.024 |
| B | 0.095 | 0.090 | 0.094 | 0.094 | 0.12 | 0.035 |
| C | 0.096 | 0.092 | 0.095 | 0.095 | 0.12 | 0.029 |
| D | 0.098 | 0.088 | 0.098 | 0.090 | 0.12 | 0.026 |
| E | 0.099 | 0.089 | 0.097 | 0.094 | 0.10 | 0.032 |

**Table 2 Comparison of Computation Time (sec)**

| SET | ARMA (2,0) | ARMA (2,1) | ARMA (2,2) | ARMA (3,0) | ARMA (3,1) | KF |
|---|---|---|---|---|---|---|
| A | 20 | 25 | 26 | 27 | 29 | 0.54 |
| B | 18 | 24 | 25 | 25 | 24 | 0.53 |
| C | 15 | 24 | 24 | 25 | 22 | 0.50 |
| D | 12 | 23 | 24 | 23 | 21 | 0.49 |
| E | 12 | 20 | 21 | 21 | 20 | 0.48 |